\documentclass[a4paper,11pt]{article}
\usepackage{pos}
\usepackage{amsmath}
\usepackage{bm}
\usepackage[normalem]{ulem} 
\usepackage{color}\definecolor{RED}{rgb}{1,0,0}\definecolor{BLUE}{rgb}{0,0,1} 
\usepackage{changebar}
\usepackage{ulem}
\usepackage{color}
\usepackage{lineno}

\title{Non-identical particle femtoscopy in Pb$-$Pb collisions at 5.02 TeV with ALICE}

\author{Pritam Chakraborty}

\affiliation{Department of Physics, Indian Institute of Technology Bombay,\\
  Main Gate Rd, IIT Area, Powai, Mumbai, Maharashtra, India,  Pin: 400076}
\emailAdd{pritam.chakraborty@cern.ch}
\note{For the ALICE Collaboration.}

\abstract{ The non-identical particle femtoscopy is a technique that is developed to estimate the dimension of a particle-emitting medium as well as the average pair-emission asymmetry between the particles using two-particle correlation functions. The analysis of femtoscopic correlations for all possible charged combinations of pion-kaon pairs in Pb$-$Pb collisions at $\sqrt{s_{\mathrm{NN}}}=$ 5.02 TeV with ALICE at the LHC is presented in this article. The results are extracted for different centrality classes of Pb$-$Pb collisions using the spherical harmonics representations of the correlation functions, $C^{\rm 0}_{\rm 0}$ and $Re C^{\rm 1}_{\rm 1}$. The evolution of the femtoscopic parameters with multiplicity and collectivity of the system is discussed extensively. The femtoscopic parameters are observed to increase with the multiplicity. The non-zero values of pair-emission asymmetry and the decrease of source-size with increasing pair-transverse momentum indicate the presence of collectivity in the system.}

\FullConference{%
  41st International Conference on High Energy physics - ICHEP2022\\
  6-13 July, 2022\\
  Bologna, Italy
}

\begin{document}
\maketitle

\section{Introduction}

The femtoscopic analysis of the two-particle correlation functions has been used since a long time to estimate the spatio-temporal dimension of the particle-emitting source, produced in heavy-ion collisions \cite{intro1}. Non-identical particle femtoscopy provides a scope to measure the asymmetry between the average emission points of two particle species from a medium along with its size. In this report, the analysis of femtoscopic correlations between all charged combinations of pion-kaon pairs in Pb--Pb collisions at $\sqrt{s_{\rm NN}}=$ 5.02 TeV with ALICE at the LHC is discussed.

\section{Formalism of non-identical particle femtoscopy} \label{theory}

The two-particle femtoscopic correlation function can be expressed as \cite{adam_primary}:
\begin{eqnarray}
	C({\bm p}_{\rm 1}, {\bm p}_{\rm 2}) = \frac{W_2({\bm p}_{\rm 1}, {\bm p}_{\rm 2})}{W_1({\bm p}_{\rm 1})W_1({\bm p}_{\rm 2})} \label{eq1}  
\end{eqnarray}
where, $W_2({\bm p}_{\rm 1},{\bm p}_{\rm 2})$ is the probability of detecting particle 1 with momentum ${\bm p}_{\rm 1}$ when particle 2 with the momentum ${\bm p}_{\rm 2}$ is detected, while $W_1({\bm p})$ is the probability of detecting particles with momentum ${\bm p}$. In case of absence of any correlation, $C({\bm p}_{\rm 1}, {\bm p}_{\rm 2})$ is equal to one. 

The dimensions of a system are represented by $out$, $side$ and $long$ axes, where $out$ is along the pair-transverse momentum ($k_{\rm T}$), $long$ is along the beam direction and $side$ is perpendicular to the other two axes. In the Pair Rest Frame (PRF), the total momentum of the pair is zero. The correlation between two particles having relative separation $\bm r^*$ and relative pair momentum $\bm k^*$ ( (*) corresponds to PRF) can be represented as the convolution of the source function, i.e. the emission probability of the pair, S($\bm r^*,\bm k^*$), and the interaction between the particles, expressed by the wave function $\Psi(\bm k^*,\bm r^*)$, as given in Eq.~\ref{eq:psi} \cite{adam_primary}.
\begin{equation}
	C(\bm k^*) = \int S(\bm k^*, \bm r^*)|\Psi(\bm k^*,\bm r^*)|^2d^3r^* \label{eq:psi}
\end{equation}
It is assumed that the source function of a non-identical particle pair is a three-dimensional Gaussian function with sizes $R_{\rm out}$, $R_{\rm side}$ and $R_{\rm long}$ in the $out$, $side$ and $long$ directions, respectively and the pair-emission asymmetry, $\mu_{\rm out}$, along the $out$ direction as given in Eq.~\ref{eq:sourcefcn} \cite{adam_primary}. 
\begin{equation}
	S(\mathbf{r})=exp\left(-\frac{(r_{\rm out}-\mu_{\rm out})^2}{2R^2_{\rm out}}-\frac{r^2_{\rm side}}{2R^2_{\rm side}}-\frac{r^2_{\rm long}}{2R^2_{\rm long}}\right)  \label{eq:sourcefcn}
\end{equation}
The parameter $\mu_{\rm out}$ is estimated as the difference between the average emission point of the light and heavy particles i.e. $\mu_{\rm out} =<x^{\rm light}_{\rm out} - x^{\rm heavy}_{\rm out}>$ and it arises only when the collective and the random thermal velocities of the particles are comparable \cite{adam_primary}. The wave function $\Psi(\bm k^*,\bm r^*)$ includes the strong and relatively more dominating Coulomb interaction between the particles in a pair. It can be computed analytically.

The experimental correlation function can be constructed by taking the ratio of $A(k^*)$ to $B(k^*)$, which are the three-dimensional particle-pair distributions selected from same and mixed events, respectively as given in Eq.~\ref{eq:C}.
\begin{equation}
	C(\bm k^*) = A(\bm k^*)/B(\bm k^*) \label{eq:C}
\end{equation}
However, the correlation functions can also be constructed in terms of spherical harmonics components in the following way \cite{adam_sh}: the $\bm{k^*}$ i.e. $k^*_{\rm out}$, $k^*_{\rm side}$ and $k^*_{\rm long}$ of the pairs from the respective same and mixed events can be decomposed into $|\bm k^*|$, $\theta_{\bm k^*}$ and $\phi_{\bm k^*}$, which are used to calculate the corresponding spherical harmonics i.e. $Y_{\rm lm}(\theta_{\bm k^*},\phi_{\bm k^*})$ (see Eq.~\ref{eq:Ylm}). The respective numerator and denominator are filled according to the calculated weights according to Eq.~\ref{eq:num_sh}-\ref{eq:den_sh} \cite{adam_sh}.
\begin{eqnarray}
	A(\bm k^*)=\sqrt{4\pi}\sum_{\rm l,m}A_{\rm lm}(\bm k^*)Y_{\rm lm}(\theta_{\bm k^*}, \phi_{\bm k^*}) \label{eq:num_sh} \\
	B(\bm k^*)=\sqrt{4\pi}\sum_{\rm l,m}B_{\rm lm}(\bm k^*)Y_{\rm lm}(\theta_{\bm k^*}, \phi_{\bm k^*}) \label{eq:den_sh}
\end{eqnarray}
where,
\begin{eqnarray}
	Y_{\rm lm}(\theta,\phi)=(-1)^{\rm m}\left[\frac{(2\rm {l+1})(\rm  {l-m})!}{4{\rm \pi(l+m)}!}\right]P_{\rm l}^{\rm m}(cos\theta)e^{i\rm{m}\phi},  \label{eq:Ylm}
\end{eqnarray}
Finally, the spherical harmonics components of the correlation function are estimated using the following relation (Eq.~\ref{eq:C_SH}). 
\begin{equation}
	A(\bm k^*)=C(\bm k^*)B(\bm k^*) \label{eq:C_SH}
\end{equation}
The advantages of using SH representations are the following: most terms vanish due to intrinsic symmetries, only few lower harmonics of $l$ and $m$ and relatively few statistics are needed to extract relevant femtoscopic parameters. Analysing only $C_{\rm 0}^{\rm 0}$ and $Re C_{\rm 1}^{\rm 1}$, one can estimate the $R_{\rm out}$ and $\mu_{\rm out}$, respectively.

The femtoscopic parameters can be extracted by parameterising the experimental correlation functions with the source function and the pair-interaction using Eq.~\ref{eq:psi}. In this analysis, the $C_{\rm 0}^{\rm 0}$ and $Re C_{\rm 1}^{\rm 1}$ components of the pion-kaon correlation functions are used to extract the femtoscopic parameters of the source.

\section{Analysis details} \label{ana}
\begin{figure}[!h]
				\centering
				{\includegraphics[width=0.65\textwidth]{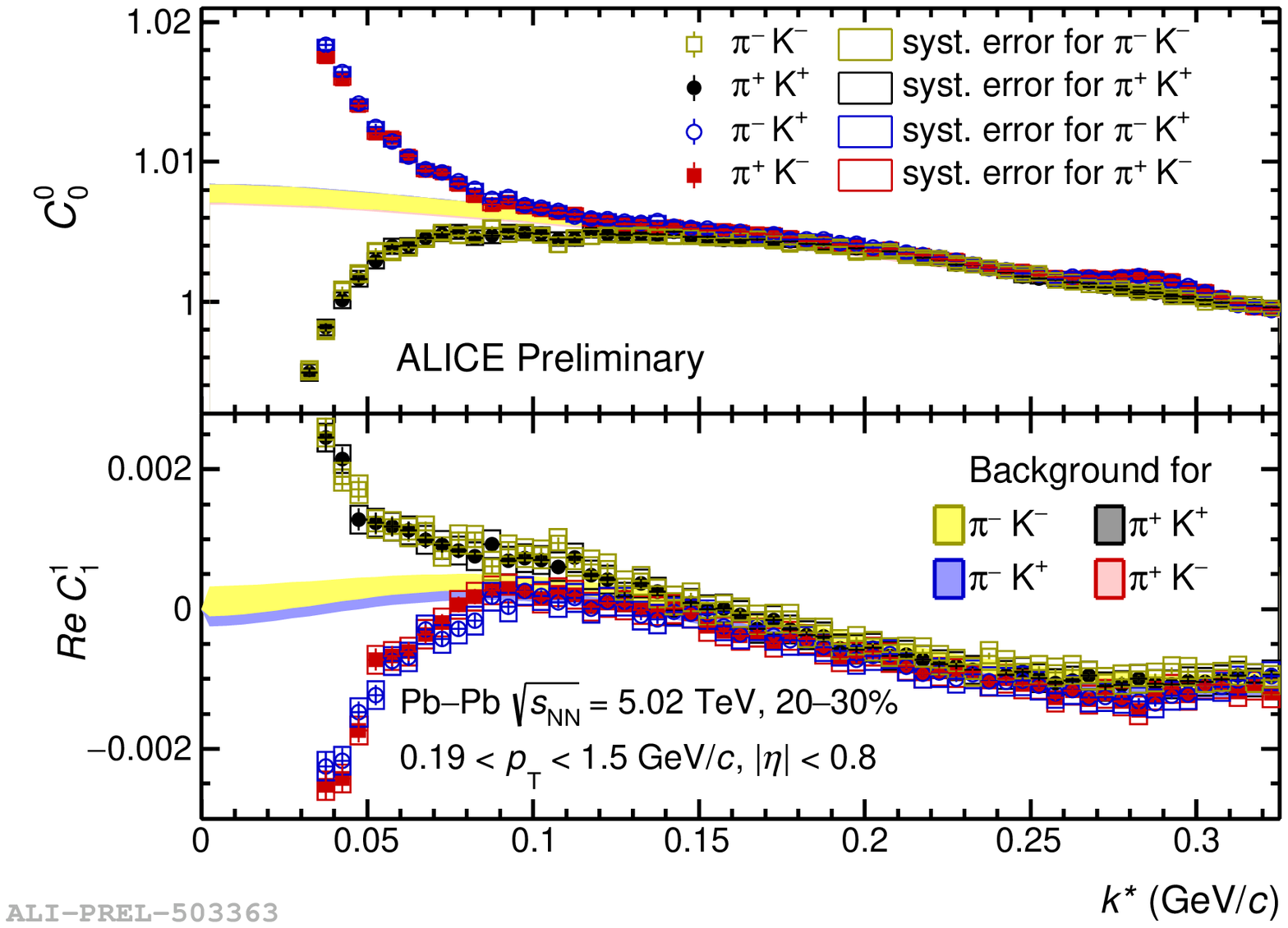}}
				{\includegraphics[width=0.65\textwidth]{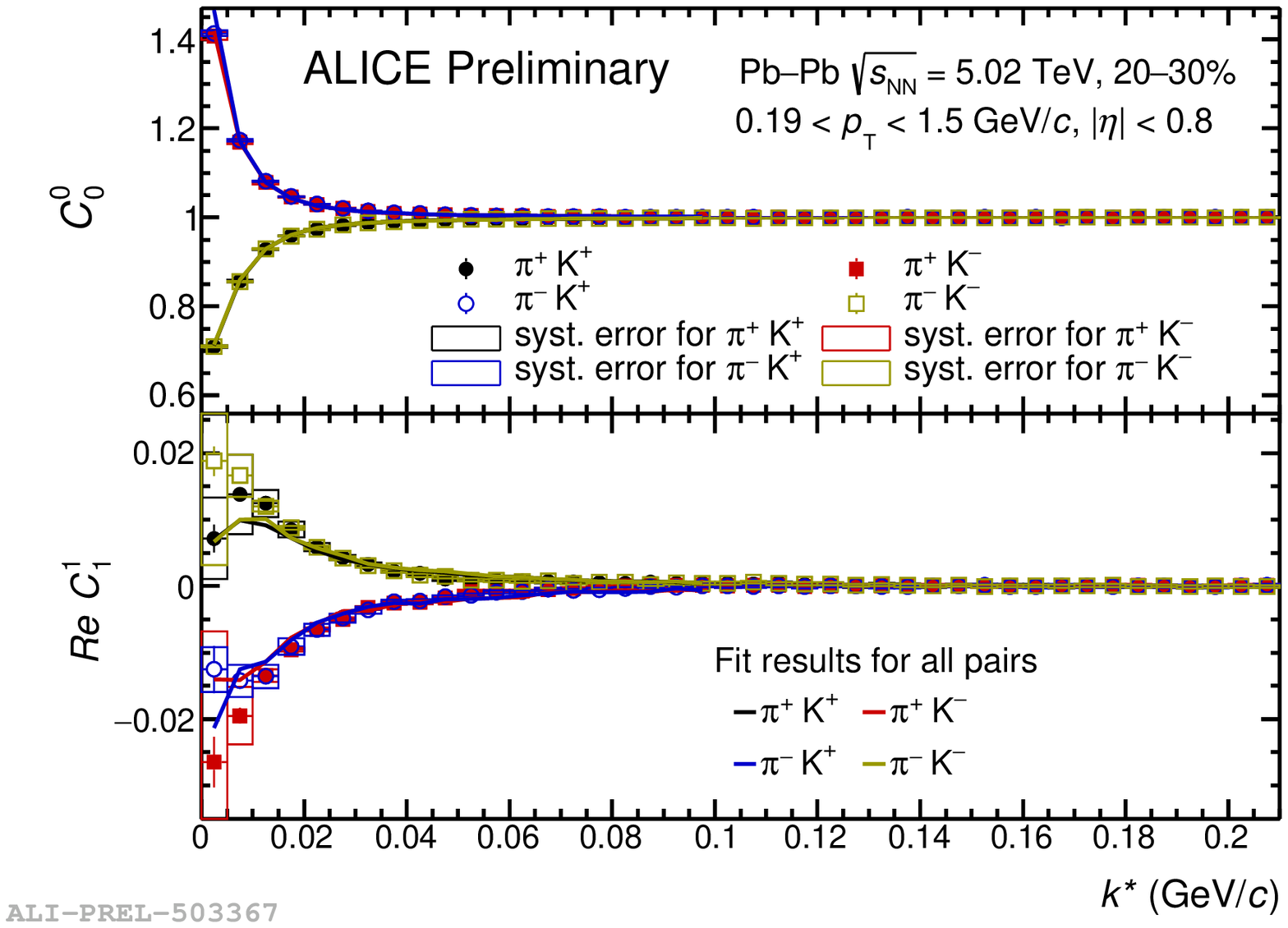}}
				\caption{Correlation functions for pion-kaon pairs with $6^{th}$ order polynomial function as the non-femtoscopic background (upper) and background-minimised femtoscopic correlation functions with the fit results, estimated using CorrFit software (bottom) for 20-30\% central Pb$-$Pb collisions at $\sqrt{s_{\rm NN}} = 5.02$ TeV}
				\label{CF_fit}
				\end{figure}

The spherical harmonic components of the correlation functions of all charged combinations of pion-kaon pairs were studied for 0-5\% to 40-50\% centrality classes in Pb$-$Pb collisions at $\sqrt{s_{\rm NN}}=5.02$ TeV with ALICE. The sub-detectors mainly used in this analysis are the ITS or primary vertex reconstruction, the TPC and the TOF for tracking and particle identification and, finally, the V0 for trigger selection and centrality estimation. The events having the z-position of the primary vertex ($\rm {z_{vtx}}$) within 7.0 cm of the nominal interaction point were selected. The pions and kaons were selected within the kinematic range of $|\eta|$<0.8 and 0.19 <$p_{\rm T} (\rm{GeV}/c)$< 1.5. The $C_{\rm 0}^{\rm 0}$ and $Re C_{\rm 1}^{\rm 1}$ components for all charged combinations of pion-kaon pairs are shown in the upper panel of Fig.~\ref{CF_fit}. It is observed that the $C_{\rm 0}^{\rm 0}$ for pion-kaon pairs of same- and opposite-sign become lower and higher than unity, respectively in the lower $k^*$ region due to the strong interaction and the Coulomb interaction. Similarly, the $Re C_{\rm 1}^{\rm 1}$ values deviate from zero in the lower $k^*$ region, indicating the possible presence of emission asymmetry along the $out$ direction between the pion and kaon in a pair. However, the $C_{\rm 0}^{\rm 0}$ is always larger than unity for opposite-sign pairs, while for same-sign pairs, it is less than unity at $k^*$< 0.05 GeV/$c$. Also, the $Re C_{\rm 1}^{\rm 1}$ is almost negative for opposite-sign pairs, while for same-sign pairs, it becomes positive for $k^*$< 0.15 GeV/$c$ and then turns negative. 
Moreover, the femtoscopic correlations should vanish in the high $k^*$ region, therefore the $C_{\rm 0}^{\rm 0}$ and $Re C_{\rm 1}^{\rm 1}$ should be flat. However, a distinct shape of the correlation functions is observed in this region, indicating the presence of non-femtoscopic background in the measured correlation function (given in Eq.~\ref{eq:back1}) due to residual correlation functions, elliptic flow, resonance decays etc. The background is present in the whole $k^*$ region and it uplifts the femtosocpic correlation functions in the lower $k^*$ region.  The experimental correlation function, $C^{ij}_{exp}$, can therefore be expressed as follows:
\begin{eqnarray}
	C^{ij}_{exp}=C^{ij}_{real} + B^{ij},  \label{eq:back1}
 \end{eqnarray}
 where, $C^{ij}_{real}$ is the real femtoscopic correlation, $B^{ij}$ is the background and $i,j$ are positively and negatively charged pair combinations. The background is similar for all four charge-pair combinations of pions and kaons and is estimated to be a polynomial function of $6^{th}$ order as shown as  bands in the upper panel of Fig.~\ref{CF_fit} \cite{adam_pik_back}. The background-minimised correlation functions are fitted using a package, called $CorrFit$ \cite{corrfit}, assuming $R_{\rm side} = R_{\rm out}$ and $R_{\rm long} = 1.3R_{\rm out}$, on the basis of the results from pion-pion 3D femtoscopic analysis in ALICE \cite{alice_pi}. The fit results are shown in the bottom panel of Fig.~\ref{CF_fit}.

\section{Results and discussions} \label{result}
\begin{figure}[!h]
				\centering
				{\includegraphics[width=0.68\textwidth]{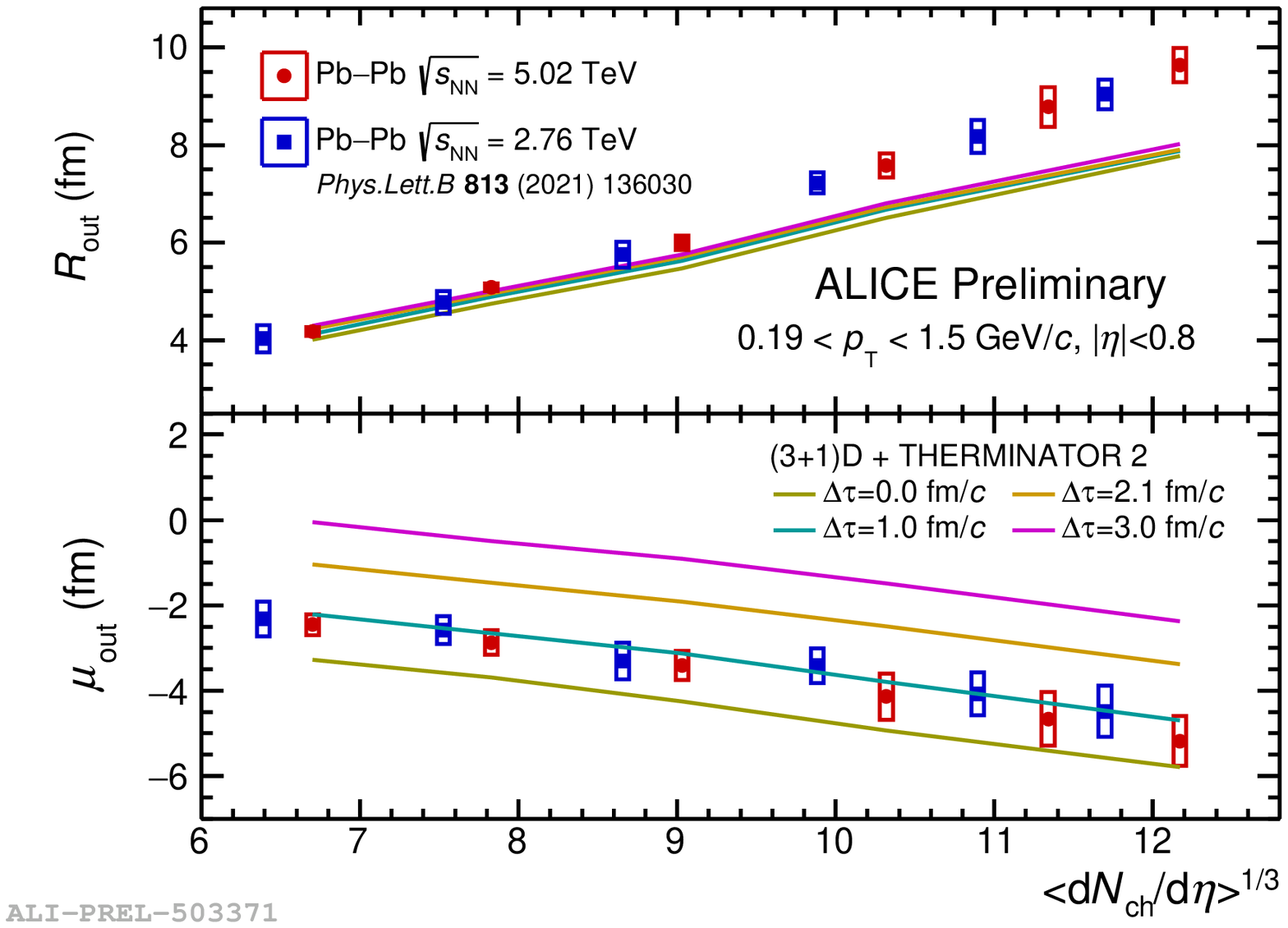}}
				{\includegraphics[width=0.68\textwidth]{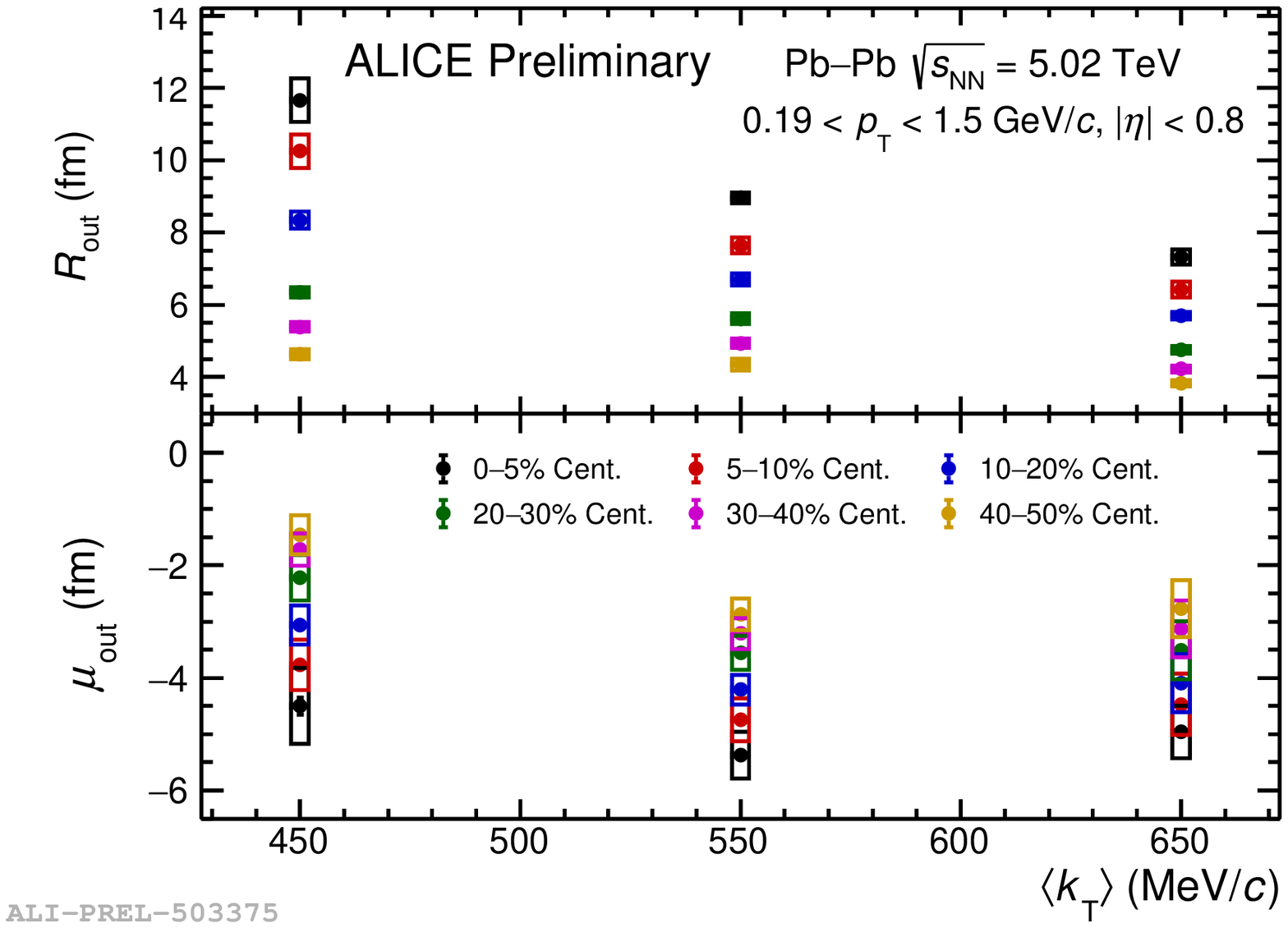}} 
				\caption{$R_{\rm {out}}$ and $\mu_{\rm {out}}$ as the function of $\langle {\rm d}N_{ch}/{\rm d}{\eta}\rangle^{1/3}$ with predictions from (3+1)D + THERMINATOR 2 model calculations (upper) and $R_{\rm {out}}$ and $\mu_{\rm {out}}$ as the function of $\langle k_{\rm T}\rangle$ and centrality classes (bottom) in Pb$-$Pb collisions at $\sqrt{s_{\rm NN}} = 5.02$ TeV}
				\label{R_mu}
			\end{figure}

The $R_{\rm {out}}$ and $\mu_{\rm {out}}$ obtained by fitting the correlation functions of all charged combinations of pion-kaon pairs from 0-5\% to 40-50\% central Pb$-$Pb collisions at $\sqrt{s_{\rm NN}}=5.02$ TeV, are shown as the function of $\langle {\rm d}N_{ch}/{\rm d}{\eta}\rangle^{1/3}$ in the upper panel of Fig.~\ref{R_mu}. The $R_{\rm {out}}$ increases with multiplicity which is expected as the system size increases with increasing number of participants. The results were compared with the predictions from (3 + 1)D viscous hydrodynamics coupled to THERMINATOR 2 model calculations, introducing different values of additional delay in kaon-emission time ($\Delta\tau$). The predictions agree well with the measured values in peripheral events, but underestimates them in central and mid-central events. The observation of non-zero $\mu_{\rm {out}}$ values indicates the existence of radial flow in the system. Moreover, the $\mu_{\rm {out}}$ is found to be negative in every multiplicity class, which suggests that the pions are emitted closer to the center of the system than kaons. It is also observed that the measured values of $\mu_{\rm {out}}$ lie between the lines corresponding to the prediction with $\Delta\tau$=0 fm/$c$ and $\Delta\tau$=1 fm/$c$. This suggests that a hadronic-rescattering phase may also contribute to the development of these correlations as a function of $\langle {\rm d}N_{ch}/{\rm d}{\eta}\rangle^{1/3}$. Finally, the results are compared with the pion-kaon femtoscopic analysis in Pb$-$Pb collisions at $\sqrt{s_{\rm NN}}=$ 2.76 TeV and no energy dependence within the current uncertainties is observed \cite{ashu}. 

The femtoscopic parameters are also been shown as a function of <$\langle k_{\rm T}\rangle$> in different centrality classes in the bottom panel of Fig.~\ref{R_mu}. The $R_{\rm {out}}$ decreases with increasing $\langle k_{\rm T}\rangle$ in all centralities. This is expected because the region of homogeneity decreases when the two particles are emitted in similar direction with high $p_{\rm T}$ values. The $\mu_{\rm {out}}$ value is observed to be lowest in the $k_{\rm T}$ range: 400-500 MeV/\textit{c} in every centrality class. However, it needs to be studied in more $k_{\rm T}$ bins to understand how it is affected by collectivity.

\section{Summary}
The correlation functions of charged pion-kaon pairs in Pb$-$Pb collision at $\sqrt{s_{\rm NN}}=$5.02 TeV with ALICE experiment were analysed. It is observed that the source size increases with the number of participants and decreases with increasing pair-transverse momentum due to the development of collective effects. The pair-emission asymmetry is estimated to be negative, implying that pions are emitted closer to the center of the system than kaons. Moreover, the prediction from Monte Carlo model calculations suggests the possible presence of hadronic rescattering phase in the system. Besides, no beam-energy dependence of the femtoscopic parameters has been found so far.


\begin{thebibliography}{99}
\bibitem{intro1}{S.~Acharya \textit{et al.} [ALICE], Phys. Rev. C \textbf{96}, 064613 (2017)}
\bibitem{adam_primary} A.~Kisiel, Phys. Rev. C, {\bf 81}, 064906 (2010) 
\bibitem{adam_pik_back}A.~Kisiel, Acta Phys. Polon. B \textbf{48}, 717 (2017)	
\bibitem{ashu}S.~Acharya \textit{et al.} [ALICE], Phys. Lett. B \textbf{813}, 136030 (2021)
\bibitem{adam_sh}A.~Kisiel, and D.~A.~Brown, Phys. Rev. C {\bf 80}, 064911 (2009)
\bibitem{alice_pi}J.~Adam \textit{et al.} (ALICE Collaboration) Phys. Rev. C {\bf 93}, 024905 (2016)
\bibitem{hotqcd}S.~K.~Das \textit{et el.},	arXiv:2208.13440
\bibitem{corrfit}A.~Kisiel, Nukleonika, {\bf 49}(suppl2), S81-S83 (2004)
\end{thebibliography}
\end{document}